\documentclass[aps,pra,preprint,showpacs,floatfix]{revtex4}
\usepackage{bm}
\usepackage{graphicx} % for PDF output
\usepackage{amsmath}
\usepackage{subfigure}
\begin{document}
\title{Differential Scattering and Rethermalization in Ultracold Dipolar Gases}
\author{John L. Bohn}
\author{Deborah S. Jin}
\affiliation{JILA, NIST, and Department of Physics, University of Colorado, Boulder, Colorado 80309, USA}

\date{\today}

\begin{abstract}
Analytic expressions for the differential cross sections of ultracold atoms and molecules that scatter primarily
due to dipolar interactions are derived within the first Born approximation,  and are shown to agree with the partial wave expansion.  These cross sections are applied to the problem of cross-dimensional rethermalization.  Strikingly, the rate of rethermalization can vary by as much as a factor of two, depending on the orientation of polarization of the dipoles.  Thus the anisotropic dipole-dipole interaction can have a profound effect even on the behavior of a nondegenerate ultracold gas.
\end{abstract}


\maketitle

\section{Introduction}

It is commonly appreciated that the ultracold regime of collisions occurs when only a small number of partial waves contributes to scattering.  This is a sensible criterion for atoms interacting via van der Waals forces, where the Wigner threshold laws dictate that scattering phase shifts scale linearly with wave number $k$ for $s$-wave collisions, but diminish as higher powers of $k$ for higher partial waves \cite{Sadeghpour00_JPB}.  For atoms or molecules scattering via dipolar forces, however, this criterion fails, since the threshold phase shifts for {\it all} nonzero partial waves scale linearly in $k$  \cite{Shakeshaft72_JPB}. Thus in principle a large number of partial waves may be necessary to describe scattering of polar species even in the zero-temperature limit.  

It is therefore worthwhile to describe the scattering of polar species directly in coordinate space, rather than in terms of angular momentum quantum numbers.  In this article we derive the relevant differential cross sections within the Born approximation.  This approach directly reveals the angular structure of the scattering, including the fact that, for polarized molecules, the differential cross section depends on the incident angle as well as the scattering angle.  We show that the partial wave series version of the Born approximation formally converges to the correct angular distribution, but that the convergence may be extremely slow.

We illustrate the influence of the anisotropy of scattering by using it to calculate the rate of rethermalization in a classical gas that has been brought out of thermal equilibrium.  While dipole interactions are now well-known to influence the behavior of quantum degenerate Bose gases \cite{Lahaye08_PRL}, we illustrate here a significant effect of the anisotropy of collision in a thermal (but still ultracold) gas.  These results  have implications for collisions and especially for evaporative cooling of polar molecules or highly magnetic atoms such as dysprosium \cite{Lu11_PRL,Lu12_PRL} and erbium \cite{Aikawa12_PRL,Aikawa13_preprint}.

%%%%%%%%%%%%%%%%%%%%%%%%%%%%%%%%%%%%%%%%%%%%

\section{Cross Sections}

For the sake of concreteness, we will speak of polar molecules polarized in an electric field ${\cal E}$, although the results apply equally to magnetic atoms polarized in a magnetic field.  These molecules interact via the dipole potential (in CGS units)
\begin{eqnarray}
\label{dipole_potential}
V_d({\bf r}) &=& \frac{{\bf d}_1 \cdot {\bf d}_2 - 3 ( {\hat r} \cdot
{\bf d}_1) ({\hat r} \cdot {\bf d}_2)}{r^3} \nonumber \\
&=& -\frac{2d^2}{r^3} C_{20}
(\theta,\phi).
\end{eqnarray}
Here ${\bf d}_{1,2}$ are the molecular dipole moments, ${\bf r}$ is the
relative coordinate joining them, and $C_{20} = (3 \cos^2 \theta -1)/2$ is
a reduced spherical harmonic \cite{BS}.  Writing the interaction in this way assumes that the electric-field direction coincides with the laboratory $z$-axis.  This orientation simplifies the derivation of the cross section, but subsequently the cross section will be cast in a coordinate-independent form.  
In the final expression on the right of (\ref{dipole_potential}) we employ a single dipole moment $d$
as if the molecules were the same.  More generally, the replacement
$d^2 \rightarrow d_1 d_2$ can be made, if the dipoles are different.  

The Hamiltonian describing scattering of the molecules, in their center-of-mass frame, is
\begin{eqnarray}
\label{Hamiltonian}
H = - \frac{ \hbar^2 }{ 2 \mu } {\vec \nabla }^2
+V_d + V_{sr},
\end{eqnarray}
where, as usual, $\mu$ stands for the reduced mass of the collision partners. In (\ref{Hamiltonian}) we allow for an additional short-range interaction $V_{sr}$ that can describe an $s$-wave phase shift not already accounted for by the dipole potential.  It is convenient to convert the scattering problem into dipole units, where
the natural length and energy scales are, respectively,
\begin{eqnarray}
a_d = \frac{ \mu d^2 }{ \hbar^2}, \;\;\;\;
E_d = \frac{ \hbar^6 }{ \mu^3 d^4 }.
\end{eqnarray}
Cast in these units, the Schr\"{o}dinger equation reads
\begin{eqnarray}
\left[ - \frac{ 1 }{ 2 } \vec{ \nabla}^2 - \frac{ 2C_{20} }{ r^3 } + V_{sr} \right] \psi = \epsilon \psi,
\end{eqnarray}
where $\epsilon = (\hbar^2 k^2 / 2 \mu)/ E_d$ is the collision energy in reduced units.

%%%%%%%%%%%%%%%%%%%%%%%%%%%%%

\subsection{Scattering Amplitudes}

Scattering in three dimensions is characterized by a differential
cross section that relates the wave vector ${\bf k}$ of the molecules'
initial approach to one another, to the wave vector ${\bf k}^{\prime}$
into which they scatter:
\begin{eqnarray}
\frac{ d \sigma }{ d \Omega_{{\bf k}^{\prime}} }({\bf k}^{\prime},{\bf k})
= |f({\bf k}^{\prime},{\bf k})|^2,
\end{eqnarray}
written, as is conventional, in terms of a scattering amplitude $f$.
The total scattering cross section is then
\begin{eqnarray}
\sigma ({\bf k}) = \int d \Omega_{{\bf k}^{\prime}}
\frac{ d \sigma }{ d \Omega_{{\bf k}^{\prime}} }({\bf k}^{\prime},{\bf k}),
\end{eqnarray}
which for dipoles can depend on the orientation of the incident
direction ${\bf k}$ with respect to the polarization axis.

In the case of indistinguishable bosons or fermions, symmetry
places constraints on the scattering amplitudes.  Specifically,
the amplitude for scattering from a given incident direction ${\bf k}$
must by properly symmetrized with respect to two opposing
scattering directions $\pm {\bf k}^{\prime}$,
\begin{eqnarray}
f_{B,F}({\bf k}^{\prime},{\bf k}) = \frac{ 1 }{ \sqrt{2} } \left[
f({\bf k}^{\prime},{\bf k}) \pm f(-{\bf k}^{\prime},{\bf k}) \right]
\end{eqnarray}
The resulting cross sections are then \cite{Burke99_Thesis}
\begin{eqnarray}
\label{symtot}
\sigma_{B,F} =
\int d \Omega_{{\bf k}^{\prime}}
| f_{B,F}({\bf k}^{\prime},{\bf k}) |^2.
\end{eqnarray}
We stress again that we work in reduced units.  To restore
the results to ``real'' units, one must multiply
$f$ by the dipole length $a_d$, and $\sigma$ by $a_d^2$.

%%%%%%%%%%%%%%%%%%%%%%%%%%%%%%%%%%%%%%%%%%%%

\subsection{Cross Section Formulas}

It has been argued previously that scattering of dipolar
particles at ultralow temperatures is well approximated within the
first-order Born approximation
\cite{Marinescu98_PRL,Hensler03_APB,Derevianko03_PRA,Kanjilal07_PRA,Ticknor08_PRL,Bohn09_NJP}.
This idea, heretofore expressed in terms of a partial wave expansion,
reserves two exceptions.  First, the $s$-wave component of
scattering requires explicitly probing short-range behavior, and
cannot be reproduced within the Born approximation \cite{Wang08_NJP}.  Second, if there
are shape or other resonances, they too arise from short-range
physics and are not well-described by the Born approximation \cite{Roudnev09_PRA}.

We therefore consider here the explicitly non-resonant case.  Also, we
add an $s$-wave scattering length by hand.  The scattering amplitude
is therefore
\begin{eqnarray}
f({\bf k}^{\prime},{\bf k}) = -a +f^{(1)}({\bf k}^{\prime},{\bf k}),
\end{eqnarray}
where $a$ is the $s$-wave scattering length in units of $a_d$ (recalling that the non-dipolar scattering amplitude is $f = -a/(1+iak) \approx -a$ \cite{Landau_Lifshitz};
and $f^{(1)}$ denotes the first-order Born approximation for dipolar scattering.
This amplitude is given by \cite{Landau_book}
\begin{eqnarray}
\label{first_Born}
f^{(1)}({\bf k}^{\prime},{\bf k}) = - \frac{ 1 }{ 2\pi}
\int d^3 r  e^{-i{\bf k}^{\prime} \cdot {\bf r}} V_d({\bf r})
e^{i {\bf k} \cdot {\bf r}}.
\end{eqnarray}

As is typical for the first Born approximation, the scattering
amplitude depends only on the momentum transfer,
\begin{eqnarray}
{\bf q} = {\bf k} - {\bf k}^{\prime},
\end{eqnarray}
and not on the incoming and outgoing amplitudes separately.
To evaluate the integral (\ref{first_Born}) we expand
$e^{i{\bf q} \cdot {\bf r}}$ into spherical waves, to get
\begin{eqnarray}
f^{(1)}({\bf k}^{\prime},{\bf k}) &=& - \frac{ 1 }{ 2\pi}
\int d^3 r \left( - \frac{ 2 C_{20}({\hat r}) }{ r^3 } \right)
e^{i{\bf q} \cdot {\bf r}} \nonumber \\
&=& \frac{ 1 }{ \pi } \int d^3 r \frac{ C_{20}({\hat r}) }{ r^3 }
4 \pi \sum_{lm} i^l Y_{lm}^*({\hat q}) j_l(qr) Y_{lm}({\hat r}).
\end{eqnarray}
Inside this sum, the angular integral is
\begin{eqnarray}
\int d{\hat r} C_{20}({\hat r}) Y_{lm}({\hat r})
= \sqrt{ \frac{ 4 \pi }{ 5 } } \delta_{2l} \delta_{0m},
\end{eqnarray}
which reduces the sum to a single term.  The relevant radial integral
is then
\begin{eqnarray}
\label{radial_integral}
\int_0^{\infty} r^2 dr \frac{ j_2(qr) }{ r^3 } &=&
\lim_{b \rightarrow 0} \int_b^{\infty} dr \frac{ j_2(qr) }{ r } \nonumber \\
&=& \lim_{b \rightarrow 0} \left[
\frac{ \sin (qb) }{ (qb)^3 } - \frac{ \cos (qb) }{ (qb)^2 } \right]
\nonumber \\  &\approx & \frac{ 1 }{ 3 } - \frac{ (qb)^2 }{ 30 },
\end{eqnarray}
For future reference, we have explicitly given this integral as a limit
in terms of a small-$r$ cutoff radius $b$ in the dipole potential.
We have also, in the last line, explicitly note the next order correction
in $(qb)$. For the time being, however, we will consider the explicit limit
where the momentum and/or the cutoff proceed to zero, $qb \rightarrow 0$.

Substituting this integral, the Born approximation to the scattering
amplitude becomes
\begin{eqnarray}
f^{(1)}({\bf k}^{\prime},{\bf k}) = - \frac{ 2 }{ 3 }
\left( 3 \cos^2 \theta_q - 1 \right),
\end{eqnarray}
where $\theta_q$ is the angle between the momentum transfer ${\bf q}$
and the $z$ axis.  This form of the scattering amplitude stresses that, in a given scattering event with defined momentum transfer ${\bf q}$,
the projection of angular momentum about ${\bf q}$ is a conserved
quantity.  Note also that the Fourier transform of the dipole-dipole
interaction has been formulated previously, in the context of the
mean-field theory of dipolar Bose-Einstein condensates \cite{Santos00_PRL}.
Here we will exploit the same result specifically for two-body scattering.

The scattering amplitude can also be written in terms of the
incident and scattered wave numbers, as well as the direction
${\hat {\cal E}}$ of the electric field (taking this to be the $z$-axis),
\begin{eqnarray}
\cos^2 \theta_q &=& \frac{ q_z^2 }{ |{\bf q}|^2 } \nonumber \\
&=& \frac{ 1 }{ 2 } \frac{ ( {\hat k} \cdot {\hat {\cal E}}
- {\hat k}^{\prime} \cdot {\hat {\cal E}})^2 }
{   1 - {\hat k} \cdot {\hat k}^{\prime} }
\end{eqnarray}
This expression assumes elastic scattering,
$k^{\prime} = k$.  Doing so re-expresses the scattering
amplitude as
\begin{eqnarray}
\label{f_unsymmetrized}
f({\bf k}^{\prime},{\bf k})
= -a - \frac{ ( {\hat k} \cdot {\hat {\cal E}}
- {\hat k}^{\prime} \cdot {\hat {\cal E}})^2 }
{  1 - {\hat k} \cdot {\hat k}^{\prime} } + \frac{ 2 }{ 3 }.
\end{eqnarray}

Using the prescriptions above, this leads immediately to the
scattering amplitudes for antisymmetrized and symmetrized scattering
amplitudes
\begin{eqnarray}
\label{f_symmetrized}
f_F({\bf k}^{\prime},{\bf k}) &=& \frac{ 1 }{ \sqrt{2 } } \frac{
 4({\hat k} \cdot {\hat {\cal E}})({\hat k}^{\prime} \cdot {\hat {\cal E}})
 -2 \left[ ({\hat k} \cdot {\hat {\cal E}})^2
 + ({\hat k}^{\prime} \cdot {\hat {\cal E}})^2 \right]
 ({\hat k} \cdot {\hat k}^{\prime}) }
{ 1 - ({\hat k} \cdot {\hat k}^{\prime})^2 }
 \\
f_B({\bf k}^{\prime},{\bf k}) &=& \frac{ 1 }{ \sqrt{2} } \left[
-2a  - \frac{
 2({\hat k}\cdot {\hat {\cal E}} )^2
 +2({\hat k}^{\prime} \cdot {\hat {\cal E}})^2
 -4({\hat k}\cdot {\hat {\cal E}} )({\hat k}^{\prime} \cdot {\hat {\cal E}})
 ({\hat k} \cdot {\hat k}^{\prime}) }
{ 1 - ({\hat k} \cdot {\hat k}^{\prime})^2 } + \frac{ 4 }{ 3 } \right]
\nonumber
\end{eqnarray}
These are expressed in units of the dipoles length $a_d$, and
depend only on the directions of the momenta, not their magnitudes.
Note also that the value of $f$ for forward scattering, where
${\bf k}^{\prime} = {\bf k}$, is ambiguous; approaching
this value in the limit ${\bf k}^{\prime} \rightarrow {\bf k}$
depends on how this limit is taken. The result must always be finite, however, since it arises from the value of
$C_{20}(\theta_q,\phi_q)$ for some (admittedly ill-defined)
value of the angle $\theta_q$.  We will discuss this forward-scattering
singularity in more detail  below.

%%%%%%%%%%%%%%%%%%%%%%%%%%%%%%%%%%%%%%%%%%%%

\subsection{Dependence of scattering on incident direction}

Notwithstanding the formal ambiguity in the forward scattering
direction, the differential cross sections can  be
integrated over scattering angle to give total cross sections.
These cross sections are cylindrically symmetric around the field axis, and are therefore functions of the angle $\eta$
between the incident wave vector and the polarization axis:
\begin{eqnarray}
\sigma (\eta) = \int d \Omega_{{\bf k}^{\prime}}
\frac{ d \sigma }{ d \Omega_{{\bf k}^{\prime}} }
\end{eqnarray}
To evaluate these integrals it is convenient to choose a coordinate
system whose $z$-axis coincides with the incident direction ${\hat k}$.
The electric-field direction is then rotated into the $x$-$z$ plane of
this system, with ${\hat {\cal E}} = (-\sin \eta, 0 , \cos \eta)$.
In this coordinate system, the spherical coordinates of ${\bf k}^{\prime}$
constitute the scattering angles$(\theta_s, \phi_s)$.
Then, for example, the cross section for unsymmterized
scattering amplitude is
\begin{eqnarray}
\sigma(\eta) &=& \int_0^{2 \pi} d \phi_s \int_0^{\pi} \sin \theta_s d \theta_s
\left[ -a - \frac{ [\cos \eta (1 - \cos \theta_s)
+ \sin \eta \sin \theta_s \cos \phi_s ]^2 }{ 1 - \cos \theta_s }
+ \frac{ 2 }{ 3 } \right]^2 \nonumber \\
&=& \frac{ 2 \pi }{ 9 } \left[ 18a^2 - 3a(2-6 \cos^2 \eta) + 5
+ 6 \cos^2 \eta -3 \cos^4 \eta \right]
\end{eqnarray}
This result is cylindrically symmetric about the electric-field axis.  It is also convenient to define an angular average of the cross section, 
\begin{eqnarray}
{\bar \sigma} &=& \frac{ 1 }{ 2 } 
\int_{-1}^{+1} d(\cos \eta) \sigma(\eta) \\
&=& 4 \pi a^2 + \frac{ 64 \pi }{ 45 },
\end{eqnarray}
where the bar is meant here to denote the average over an assumed isotropic
 distribution of incident directions.
Here the first term is the usual, nondipolar, $s$-wave cross section. The second term is
the pure dipolar result, and is implicitly multiplied by the
square of the dipole length $a_d$.  Notice that this result,
$64 \pi / 45 \approx 1.117 + 3.351$, gives the sum of the even
and odd partial wave contributions, as calculated from
close-coupling calculations in Ref. \cite{Bohn09_NJP}.

\begin{figure}[tbp]
\centering
\includegraphics[width=1.0\columnwidth]{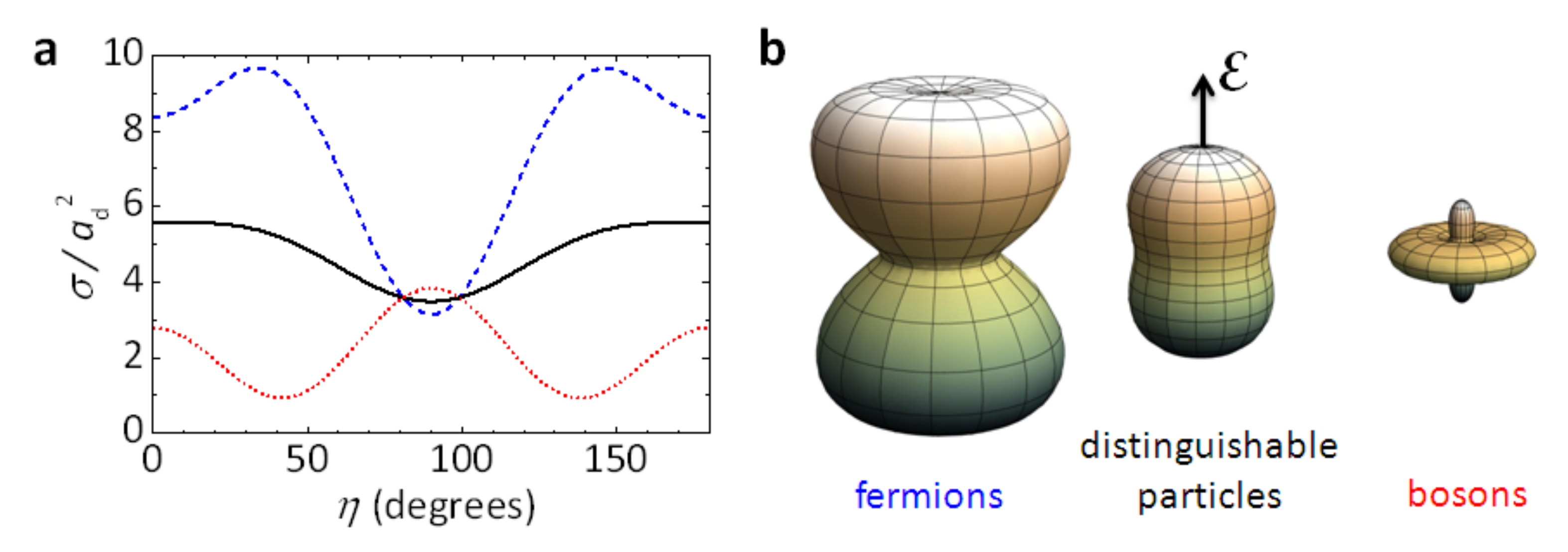}
\caption{(Color online) Total cross section ${\bar \sigma}$ for pure dipolar scattering, as a function of the angle $\eta$ between the incident direction and the polarizing electric field.  The solid black line is for distinguishable particles, while the blue dashed and red dotted lines stand for indistinguishable fermions and bosons, respectively. }
\label{sigtot}
\end{figure}

Similarly, the total cross section for indistinguishable fermions, as a function of $\eta$, is
\begin{eqnarray}
\sigma_F = \frac{ \pi }{  3 } \left[3 + 18\cos^2(\eta) -13\cos^4(\eta)  \right]
\end{eqnarray}
with angular average
\begin{eqnarray}
{\bar \sigma}_F = \frac{ 32 \pi }{ 15 };
\end{eqnarray}
and the total cross section for indistinguishable bosons is
\begin{eqnarray}
\sigma_B = \frac{ \pi }{ 9 } \left[ 72 a^2 - 24a (1-3\cos^2(\eta))
+11 - 30\cos^2(\eta)  +27\cos^4(\eta) \right],
\end{eqnarray}
with angular average
\begin{eqnarray}
{\bar \sigma}_B = 8\pi a^2 + \frac{ 32 \pi }{ 45}.
\end{eqnarray}
These total cross sections are depicted in Figure \ref{sigtot}, shown to the same scale. These figures reveal that dipolar fermions scatter more strongly than dipolar bosons.  Whereas bosons scatter most readily when they meet side-by side, fermions tend to scatter most when meeting at an angle $\eta \approx 45^{\circ}$ with respect to the field axis.

%%%%%%%%%%%%%%%%%%%%%%%%%%%%%%%%%%%%%%%%%%%%%

\section{Cross-dimensional rethermalization}

\begin{figure}[tbp]
\centering
\includegraphics[width=0.5\columnwidth]{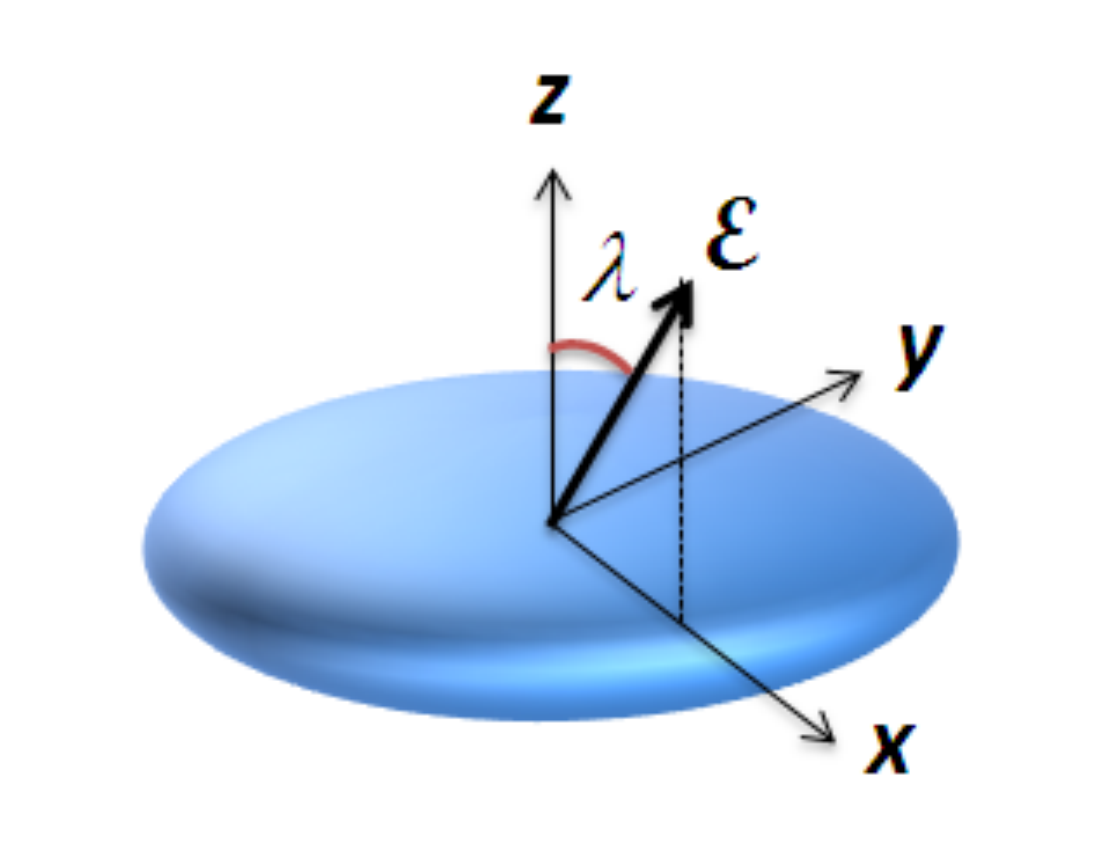}
\caption{(Color online) Geometry of a cross-dimensional rethermalization experiment.  A cylindrically symmetric cloud, with symmetry axis ${\hat z}$, is driven out of equilibrium, giving it either more or less average kinetic energy in the axial direction than the radial direction.  The rate of rethermalization is, in general, a function of the angle $\lambda$ between the symmetry axis and the electric or magnetic field that polarizes the dipoles.}
\label{retherm}
\end{figure}

An essential effect of elastic collisions for ultracold gas experiments is in thermalizing the gas, for example, during evaporative cooling.  Turning this around, measurements of rethermalization rates can be used as an experimental tool for extracting the elastic collision cross section, which is a key ingredient to understanding and designing ultracold gas experiments.  Here we point out an important effect on thermalization via dipolar elastic collisions that follows from the anisotropy of dipolar interactions. Dipolar collisions depend not only on the scattering angle ($i.e.$ the angle between ${\bf k}$ and ${\bf k}^{\prime}$) but also on the quantization direction set by the electric-field direction ${\hat {\cal E}}$.  At the same time,  traps for ultracold gases have principal axes that define a relevant coordinate system in space (which is often aligned with the Earth's gravitational field) and thermalization must include cross-dimensional thermalization, $i.e.$ keeping the energy in the different trap directions equilibrated. The axes of equilibration are shown schematically in Fig. \ref{retherm}.  A consequence of this is that the orientation of the quantization direction with respect to the trap axes, described by an angle $\lambda$,  can strongly affect thermalization rates.

\subsection{Number of collisions per thermalization}

The idea of the rethermalization  experiment is to preferentially add or remove energy along one direction of the trapped gas, then observe the energies in the different trap directions equilibrate through elastic collisions \cite{Monroe93_PRL}.  Fitting the energy in a particular trap direction versus time to an exponential, one finds a rethermalization time $\tau$, or equivalently  rethermalization rate $\gamma = 1/\tau$.  From this one extracts a rethermalization cross section defined by
\begin{eqnarray}
\gamma = n v_r \sigma_{\rm retherm},
\end{eqnarray}
in terms of the average number density $n = (1/N) \int d^3r n({\bf r})^2 $ and the mean relative speed $v_r = \sqrt{ 8 k_BT / \pi \mu }$, which are empirically determined quantities.  Here $N$ is the total number of particles, $k_B$ is Boltzmann's constant, $\mu$ is the reduced mass, and $T$ is the temperature of the gas.

The cross section $\sigma_{\rm retherm}$ extracted from rethermalization data is not, in general, identical to the average total cross section ${\bar \sigma}$, but the two are proportional:
\begin{eqnarray}
{\bar \sigma} = \alpha \sigma_{\rm retherm}.
\end{eqnarray}
The constant of proportionality $\alpha$, described by theory, allows one to extract the mean cross section from rethermalization measurements.  It can be interpreted as the ratio of a standard collision rate $n v_r {\bar \sigma}$ to the rethermalization rate $\gamma$, and hence is referred to as the ``number of collisions per rethermalization.''  For $s$-wave collisions $\alpha = 2.5$, whereas for $p$-wave collisions $\alpha = 4.17$ \cite{DeMarco99_PRL}.  The larger number for $p$-wave scattering arises from the fact the $p$-wave scattering occurs preferentially in the forward and backward scattering directions, and is thus less efficient at rethermalization than isotropic $s$-wave scattering.  For dipoles, $\alpha$ will in general be a function of the electric-field direction $\lambda$.

In the experiment, the trapping potential can be parametrically driven in a certain direction, say the $z$ direction, leading to an effective temperature $T_z$ that is greater than the effective temperature $T_y$ in the other two directions ($T_y$ could also be made the larger one, of course).  The temperatures are to be interpreted as parameters in the quasi-equilibrium phase space distribution
\begin{eqnarray}
\frac{ d^6 N }{ d^3r d^3v } ({\bf r}, {\bf v}; {\bf T}) = n({\bf r}, {\bf T}) \rho({\bf v}, {\bf T}),
\end{eqnarray}
which is given in terms of the space and velocity densities
\begin{eqnarray}
\label{MB_distribution}
n({\bf r}, {\bf T}) &=& N \Pi_{i=x,y,z}
 \left( \frac{ m \omega_i^2 }{ 2 \pi k_BT_i } \right)^{1/2} \exp \left[ - \frac{ 1 }{ 2 } \frac{ m \omega_i^2 r_i^2 }{ k_BT_i } \right]
\nonumber \\
\rho({\bf v}, {\bf T}) &=& \Pi_{i=x,y,z}  \left( \frac{ m }{ 2 \pi k_B T_i } \right)^{1/2} \exp \left[ -\frac{ 1 }{ 2 } \frac{ m v_i^2 }{ k_B T_i } \right].
\end{eqnarray}
Here the notation ${\bf T} = (T_x,T_y,T_z)$ parametrizes the mean velocities in the different directions.  This distribution is normalized so that its integral over all phase space is equal to  $N$, the total number of molecules.   

Bringing the gas out of equilibrium introduces a disparity between the mean energy per particle  in the vertical and horizontal directions, defined by
\begin{eqnarray}
\langle \chi \rangle & \equiv &  \frac{ 1 }{ N }\langle \left( \frac{ 1 }{ 2 } m \omega^2 z^2 + \frac{ 1 }{ 2 } m v_z^2 \right) 
-\left( \frac{ 1 }{ 2 } m \omega^2 y^2 + \frac{ 1 }{ 2 } m v_y^2 \right) \rangle \nonumber \\
&=&  k_B(T_z - T_y),
\end{eqnarray}
where the brackets indicate averaging over the distribution in Eqn.~(\ref{MB_distribution}).   The time evolution of this quantity, as it relaxes back to $\langle \chi \rangle = 0$,
is given by the Enskog equation \cite{Reif},
\begin{eqnarray}
\frac{ d \langle \chi \rangle }{ dt } = C(\Delta \chi),
\end{eqnarray}
in terms of a collision integral (also per particle)
\begin{eqnarray}
C(\Delta \chi) &=& \frac{ 1 }{  2N } \int d^3r_1 \int d^3v_1 \int d^3 r_2\int d^3 v_2 \frac{ d^6N }{ d^3r_1 d^3 v_1} \frac{ d^6N }{ d^3r_2 d^3 v_2}
\delta( {\bf r}_1 - {\bf r}_2)  \int d\Omega_{{\bf k}^{\prime}}
\frac{ d \sigma }{ d \Omega_{{\bf k}^{\prime}} } |{\bf v}_1 - {\bf v}_2 | \Delta \chi. \nonumber \\
\end{eqnarray}
In this expression the $\delta$ function guarantees that collisions happen locally, and the quantity that changes in a collision is the kinetic energy, so that
\begin{eqnarray}
\Delta \chi &=& \chi ({\bf v}_1^{\prime}) + \chi ({\bf v}_2^{\prime}) - \chi ({\bf v}_1) - \chi ({\bf v}_2) \nonumber \\
&=& \frac{ 1 }{ 2 } m \left( v_{1z}^{\prime 2} - v_{1y}^{\prime 2}  +v_{2z}^{\prime 2} - v_{2y}^{\prime 2} \right)
- \frac{ 1 }{ 2 } m \left( v_{1z}^2 - v_{1y}^2 + v_{2z}^2 - v_{2y}^2 \right).
\end{eqnarray}
The collision integral simplifies by separating the space and velocity dependent parts, and by representing velocities in terms of the center-of-mass velocity ${\bf V} = ({\bf v}_1 + {\bf v}_2)/2$ and the relative velocity ${\bf v}_r = {\bf v}_1 - {\bf v}_2$.  The collision integral becomes
\begin{eqnarray}
C(\Delta \chi) = \frac{  n }{ 2 } 
 \int d^3v_r \rho_r({\bf v_r},{\bf T}) \int d \Omega_{{\bf k}^{\prime}} \frac{ d \sigma }{ d \Omega_{{\bf k}^{\prime}} } v_r \Delta \chi,
 \end{eqnarray}
 where $\Delta \chi$ now reads
 \begin{eqnarray}
 \Delta \chi = \frac{ 1 }{ 2 } \mu \left( v_{r,z}^{\prime 2} - v_{r,y}^{\prime 2} \right) - \frac{ 1 }{ 2 } \mu \left( v_{r,z}^2 - v_{r,y}^2 \right)
 \end{eqnarray}
 and $  n = (1/N)\int d^3 r n({\bf r})^2$.  The velocity distribution  $\rho_r$ has the same functional form as $\rho$ in Eqn. (\ref{MB_distribution}) {\it except} that the molecular mass $m$ is replaced by the reduced mass $\mu$.

Rethermalization  proceeds at a rate characterized by
\begin{eqnarray}
\gamma = - \frac{ 1 }{ \langle \chi \rangle } \frac{ d \langle \chi \rangle }{ dt } 
\end{eqnarray}
Substituting in the above expressions and definitions, we therefore find
\begin{eqnarray}
\label{alpha}
\alpha  = \frac{ n v_r {\bar \sigma} }{ \gamma} = - \frac{ 2 k_B (T_z - T_y) v_r {\bar \sigma}   }{ \langle v_r \sigma  \Delta \chi \rangle } ,
\end{eqnarray}
using the suggestive notation
\begin{eqnarray}
\label{integral}
\langle v_r \sigma  \Delta \chi \rangle = \int d^3v_r \rho_r({\bf v}_r,{\bf T}) v_r \int d \Omega_{{\bf k}^{\prime}} \frac{ d \sigma }{ d\Omega_{{\bf k}^{\prime}} }  \Delta \chi.
\end{eqnarray}
This integral vanishes linearly as $T_z - T_y \rightarrow 0$, so that the ratio in Eqn.~(\ref{alpha}) remains well-defined in this limit.
The ratio $\alpha$ is in general a weakly dependent function of the temperature asymmetry  $T_z / T_y$.  It may however have a significant dependence on the polarization direction, owing to the anisotropy of the dipole-dipole interaction.  

%%%%%%%%%%%%%%%%%%%%%%%%%%%%%%%%%%%%%%%

\subsection{Anisotropy of rethermalization rate}

The integrals in Eqn.~(\ref{integral}) required to compute $\alpha$ can in principle be done analytically, even for the dipolar cross sections. They are, however, somewhat cumbersome, so we present numerical results here.  In this subsection we present results in the limit of small initial anisotropy, {\it i.e.}, the limit $T_z / T_y \approx 1$, and for the sake of simplicity assume that $T_x = T_y$.  As a point of reference, we note that  expression (\ref{alpha}) correctly reduces to the  results $\alpha$ = 2.5 for $s$-wave collisions and $\alpha$ = 25/6=4.17 for $p$-wave collisions.  

For collisions of dipolar particles, anisotropy of the scattering cross section implies that the rethermalization constant $\alpha$ may depend on the direction of polarization.  We therefore model a rethermalization experiment where energy transfers between the laboratory $z$ and $y$ directions, but the direction ${\hat {\cal E}}$ of the polarizing electric field is inclined at an angle $\lambda$ with respect to the $z$-axis (Fig.~\ref{retherm}). We consider separately the cases of identical fermions, identical bosons, and distinguishable particles.  

\begin{figure}[tbp]
\centering
\includegraphics[width=1.0\columnwidth]{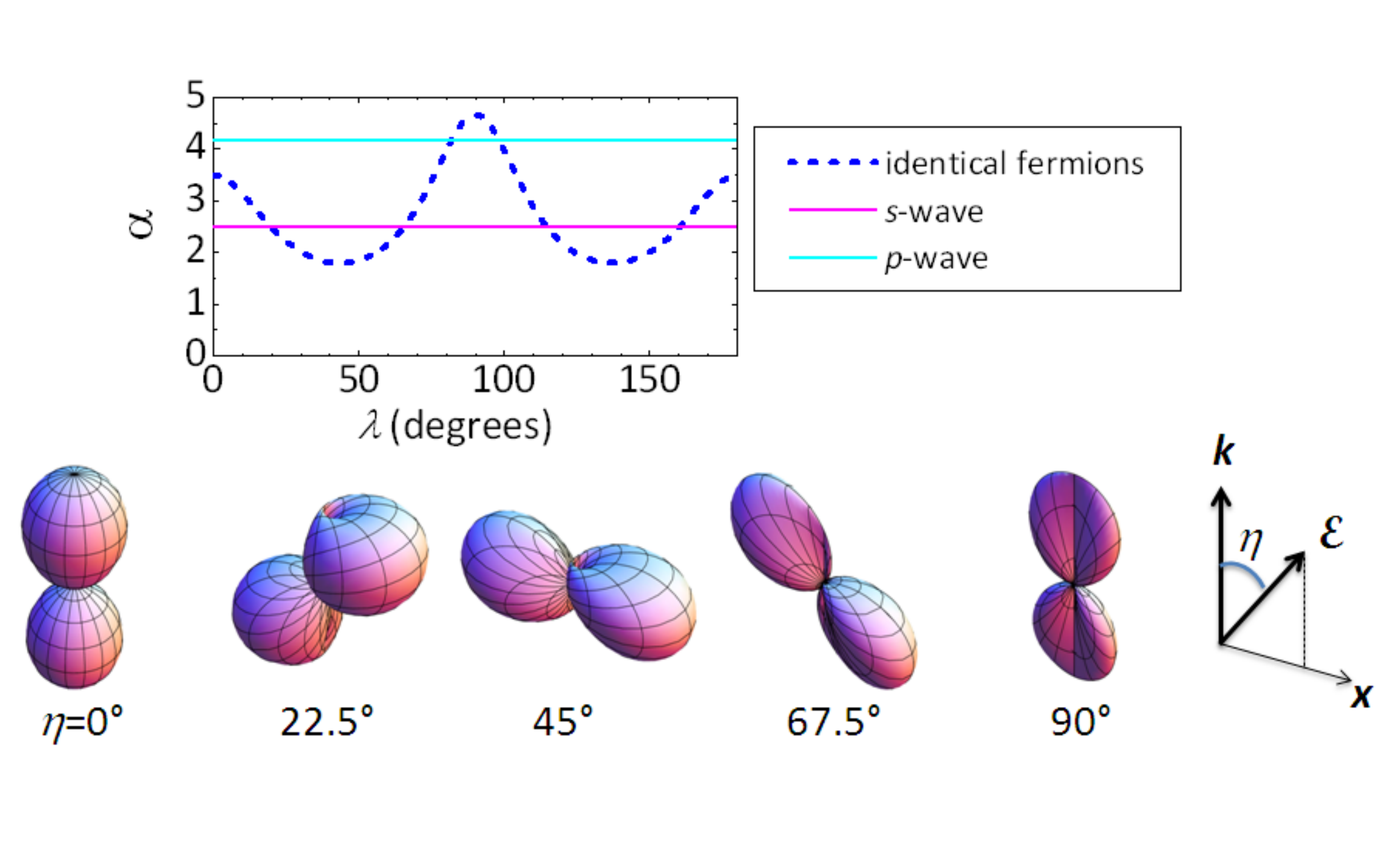}
\caption{(Color online) Upper panel: The parameter $\alpha$ that characterizes the number of collisions per rethermalization as a function of the angle $\lambda$ between the trap axis $\hat z$ and the quantization axis ${\hat {\cal E}}$.  The result for dipolar collisions of indistinguishable  fermions is shown in blue, while the cyan  and magenta lines indicate $\alpha$ for $s$-wave and $p$-wave collisions, respectively.     Lower panel: Differential cross sections for indistinguishable fermions as a function of scattering direction ${\bf k^{\prime}}$. Each plot assumes incident direction ${\bf k}$ along the vertical, and making an angle $\eta$ with respect to ${\hat {\cal E}}$.}  \label{fermion_alpha}
\end{figure}

For identical fermions, this result is presented in Fig.~\ref{fermion_alpha}.  The upper panel shows the value of $\alpha$ versus the tilt angle $\lambda$, revealing that the rethermalization rate varies by more than a factor of two.  For identical fermions, dipolar collisions are most effective at rethermalizing the gas (smallest $\alpha$)  when ${\hat {\cal E}}$ is at 45 degrees with respect to the trap axis, and least effective when $\hat{\cal{E}}$ is at 90 degrees of 0 degrees with respect to the trap axis.

To understand the behavior of $\alpha$,  it is necessary to consider the differential cross section for various values of the angle $\eta$ between the incident relative velocity and the quantization axis ${\hat {\cal E}}$.  The lower panel of Figure \ref{fermion_alpha} shows 3D surface plots of the differential cross section as a function of the direction of the outbound scattering wave vector ${\bf k}^{\prime}$.  Each such figure is drawn for a different value of the angle $\eta$ between the incident direction ${\bf k}$ (set as vertical in these diagrams) and the electric-field direction ${\hat {\cal E}}$.

Cross-dimensional rethermalization averages over many collisions with different incident angles $\eta$.  The collisions that are most efficient at cross-dimensional rethermalization require two circumstances: first, the scattering cross section must be large, and second, the scattering must change the direction of the incident velocity.  For fermions, the largest cross section occurs for collisions incident at $\eta \approx 45^{\circ}$ with respect to the electric field (Fig.~\ref{sigtot}).  Moreover, these collisions are precisely the ones that tend to preferentially scatter at large angles (center lower panel of Fig.~\ref{fermion_alpha}).  By contrast, collisions that occur with incident direction $\eta \approx 0^{\circ}$ or $90^{\circ}$ with respect to the field not only scatter less (Fig.~\ref{sigtot}), but also preferentially scatter in the forward and backward direction (Fig.~\ref{fermion_alpha}), and are not good at moving energy between dimensions.

Therefore, for an electric field tilted at an angle $\lambda = 45^{\circ}$ with respect to the laboratory $z$-axis, collisions should be relatively efficient at moving energy between radial and axial trap directions.    For an electric field oriented at $\lambda = 0^{\circ}$ or $90^{\circ}$ with respect to the $z$-axis, the most likely collisions, those with $\eta = 45^{\circ}$, occur in an orientation that is less efficient at transferring energy between laboratory radial and axial directions. This circumstance is reflected in larger values of $\alpha$ for these tilt angles.

\begin{figure}[tbp]
\centering
\includegraphics[width=1.0\columnwidth]{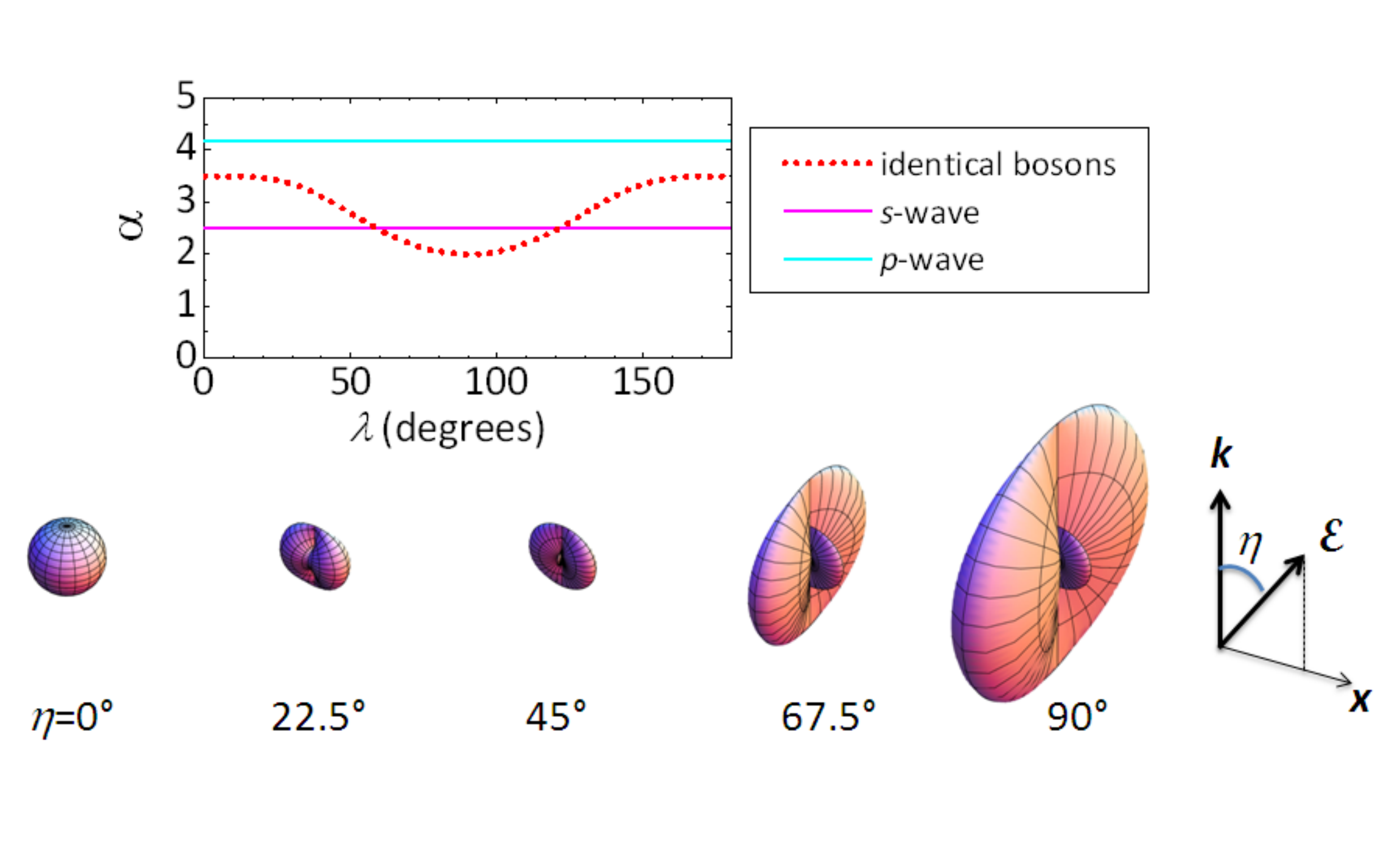}
\caption{(Color online) Upper panel: The parameter $\alpha$ that characterizes the number of collisions per rethermalization as a function of the angle $\lambda$ between the trap axis $\hat z$ and the quantization axis ${\hat {\cal E}}$.  The result for dipolar collisions of indistinguishable  bosons is shown in red, while the cyan  and magenta lines indicate $\alpha$ for $s$-wave and $p$-wave collisions, respectively.     Lower panel: Differential cross sections for indistinguishable bosons as a function of scattering direction ${\bf k^{\prime}}$. Each plot assumes incident direction ${\bf k}$ along the vertical, and making an angle $\eta$ with respect to ${\hat {\cal E}}$.}
  \label{boson_alpha}
\end{figure}

For identical bosons the situation is quite different, as shown in Fig.~\ref{boson_alpha}.  Here the tilt angle $\lambda = 90^{\circ}$ actually produces the most efficient rethermalization.  The lower panel of Fig.~\ref{boson_alpha} presents differential cross sections for various incident angles $\eta$.  Here the most likely collisions occur when $\eta=90^{\circ}$, and these collisions tend to scatter into the plane perpendicular to ${\hat {\cal E}}$.    Thus if ${\hat {\cal E}}$ is aligned perpendicularly to the $z$-axis ($\lambda =90^{\circ}$), collisions can shunt energy relatively efficiently between the axial direction and at least one radial direction in the trap.  By contrast, if ${\hat {\cal E}}$ is aligned along ${\hat z}$ ($\lambda = 0^{\circ}$), relative velocity that originates in the $x$-$y$-plane tends to remain in this plane, whereas relative velocity that originates along the $z$-axis experiences  isotropic scattering, but with a small cross section.  In either event, cross-dimensional rethermalization occurs slowly.  

\begin{figure}[tbp]
\centering
\includegraphics[width=1.0\columnwidth]{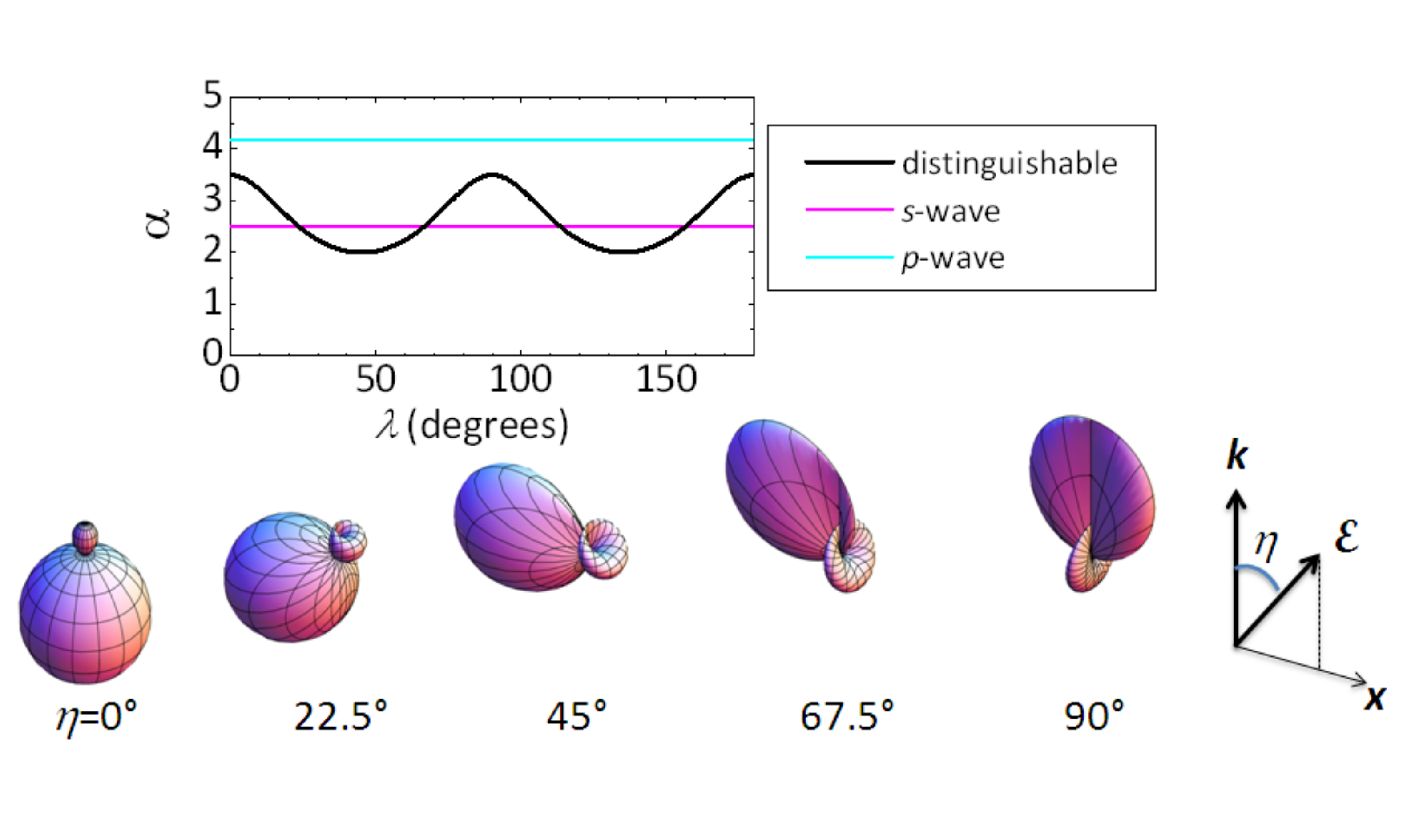}
\caption{(Color online) Upper panel: The parameter $\alpha$ that characterizes the number of collisions per rethermalization as a function of the angle $\lambda$ between the trap axis $\hat z$ and the quantization axis ${\hat {\cal E}}$.  The result for dipolar collisions of distinguishable particles is shown in black, while the cyan  and magenta lines indicate $\alpha$ for $s$-wave and $p$-wave collisions, respectively.     Lower panel: Differential cross sections for distinguishable molecules as a function of scattering direction ${\bf k^{\prime}}$. Each plot assumes incident direction ${\bf k}$ along the vertical, and making an angle $\eta$ with respect to ${\hat {\cal E}}$.}
  \label{dist_alpha}
\end{figure}

Finally, for completeness we report the angular dependence of $\alpha$ for distinguishable particles in Fig.~\ref{dist_alpha}.  Similar to fermions, in this case the most efficient rethermalization occurs  when the quantization axis ${\hat {\cal E}}$ lies at $\lambda = 45^{\circ}$ with respect to the $z$-axis, whereas, for ${\hat {\cal E}}$ either parallel or perpendicular to $z$, the scattering is primarily in the forward or backscattering direction, and does not contribute strongly to rethermalization.  

\begin{figure}
\begin{centering}
\leavevmode
\resizebox*{1.0\columnwidth}{!}{\includegraphics{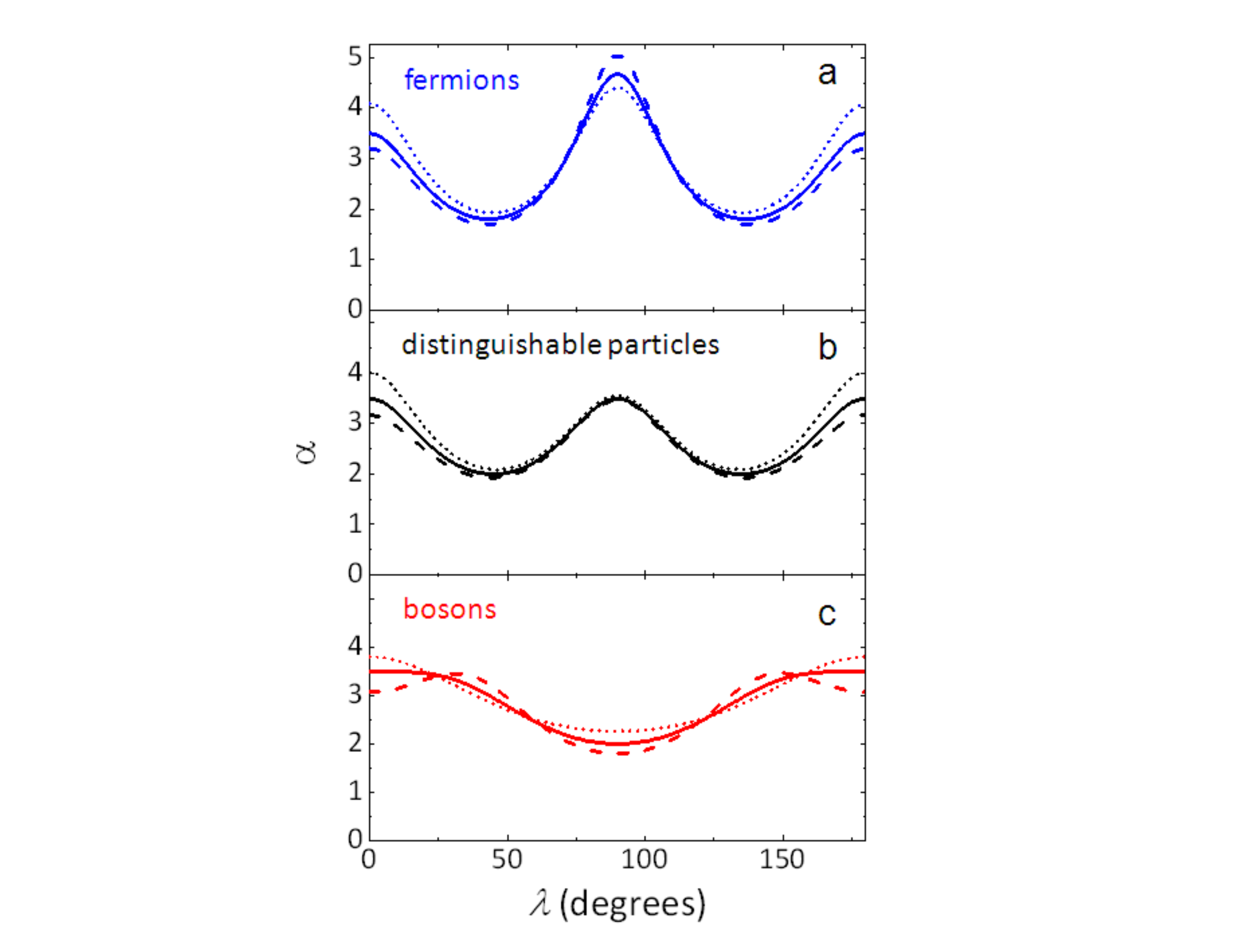}}
\end{centering}
\caption{(Color online) The dependence of $\alpha$ for cross-dimensional rethermalization on the initial temperature anisotropy for dipolar collisions of (a) identical fermions, (b) distinguishable particles, and (c) identical bosons.  The solid line shows $\alpha$ in the limit of zero anisotropy, with the dashed line corresponds to $T_z=2T_y$  and the dotted line to $T_z=0.4T_y$.} 
\label{anisotropy}
\end{figure}

%%%%%%%%%%%%%%%%%%%%%%%%%%%%%%%%%%%%%%%

\subsection{Dependence on the initial temperature imbalance}

In extracting the elastic collision cross section from measurements of cross-dimensional rethermalization, it may be useful for experimenters to know the dependence of the rethermalization rate on the initial energy (or effective temperature) imbalance. As  defined above, $\alpha$ is inversely proportional to the initial rethermalization rate, and hence is sensitive to the initial value of the imbalance $T_z/T_y$.   Fig.~\ref{anisotropy} shows $\alpha$ versus the tilt angle $\lambda$ for three different values of the initial temperature imbalance $T_z/T_y=0.4$, 1, and 2.  For comparison, for $s$- or $p$-wave collisions, $\alpha$ decreases by $3\%$ for $T_z/T_y=2$ and increases by $4\%$ for $T_z/T_y=0.4$ compared to the limit of no imbalance ($T_z/T_y=1$).  
As seen in Fig.~\ref{anisotropy}, the effect of the initial temperature imbalance can be a few times stronger than this for dipolar collisions.  This suggests that to accurately extract the elastic collision cross section from a measurement of cross-dimensional rethermalization, one would like to work in the limit of small initial energy imbalance. Alternatively, measurements taken for similar but opposite energy imbalance, namely $T_z/T_y>1$ and $T_z/T_y<1$, could be averaged. In addition, in the case of indistinguishable bosons, the shape of the $\alpha$ versus $\lambda$ curve can qualitatively change depending on the temperature imbalance.

%%%%%%%%%%%%%%%%%%%%%%%%%%%%%%%%%%%%%%%%%%%%

\section{Formal Aspects}

\subsection{Relation to the partial wave expansion}

The scattering amplitude in the first Born approximation can be expressed
as a partial wave expansion involving the spherical wave
components of both the incoming and outgoing waves:
\begin{eqnarray}
\label{partial_wave_scattering_amplitude}
f^{(1)}({\bf k}^{\prime},{\bf k}) = - \frac{ 2 \pi }{ k }
\sum_{l^{\prime}m^{\prime}lm}
i^{l^{\prime}} Y_{l^{\prime}m}^*({\hat k}^{\prime})
\langle l^{\prime} m^{\prime} |T^{\rm Born}| lm \rangle  i^{-l} Y_{lm}({\hat k}),
\end{eqnarray}
where $T^{\rm Born} = i(S^{\rm Born}-I)$ is the transition matrix 
in terms of the usual scattering matrix $S$ \cite{Mott_Massey}.
For dipolar scattering it is given by the product \cite{Kanjilal07_PRA,Bohn09_NJP}
\begin{eqnarray}
\label{Born_T}
\langle l^{\prime} m^{\prime} |T^{\rm Born}| lm \rangle
 = - (k a_d) C_{l^{\prime}l}^{(m)} \Gamma_{l^{\prime} l}
\end{eqnarray}
 in terms of the angular and radial integrals
\begin{eqnarray}
\label{angular_integral}
C_{l^{\prime}l}^{(m)} =   (-1)^m
\sqrt{ (2l^{\prime}+1)(2l+1) }
\left( \begin{array}{ccc} l^{\prime} & 2 & l \\
                          -m & 0 & m \end{array} \right)
\left( \begin{array}{ccc} l^{\prime} & 2 & l \\
                          0 & 0 & 0 \end{array} \right)
\end{eqnarray}
\begin{eqnarray}
\Gamma_{l^{\prime}l} = \left\{
\begin{array}{ll} \frac{ 4 }{ l(l+1) }, \;\;\;\;\; l^{\prime} = l \\
                  \frac{ 4 }{ 3 l(l-1) }, \;\;\;\;\; l^{\prime} = l-2 \\
                  \frac{ 4 }{ 3(l+1)(l+2) }, \;\;\; l^{\prime} = l+2
\end{array} \right.
\end{eqnarray}
(The expression for $\Gamma$ corrects typographical errors in Ref. \cite{Bohn09_NJP}).
Within the first Born approximation, then, the scattering amplitude
$f^{(1)}$ is appropriately proportional to $a_d$ and  independent of the wave number $k$.

The Born approximation (\ref{Born_T}) for the $T$-matrix is meant
to apply in the threshold limit $k a_d \ll 1$, where the scattering occurs
primarily beyond the outer classical turning point of the effective
centrifugal potential, and the magnitude of the $T$-matrix elements remains below unity.  In addition, acknowledging the possibility of $s$-wave scattering, Wang puts a second requirement on the adequacy of the Born approximation, that the ratio of the dipole length to any small-$r$ cutoff length $b$ must also be small \cite{Wang08_NJP}.  This requirement, however, is explicitly relevant only to $s$-wave scattering in the first Born approximation.

\begin{figure}[tbp]
\centering
\includegraphics[width=1.0\columnwidth]{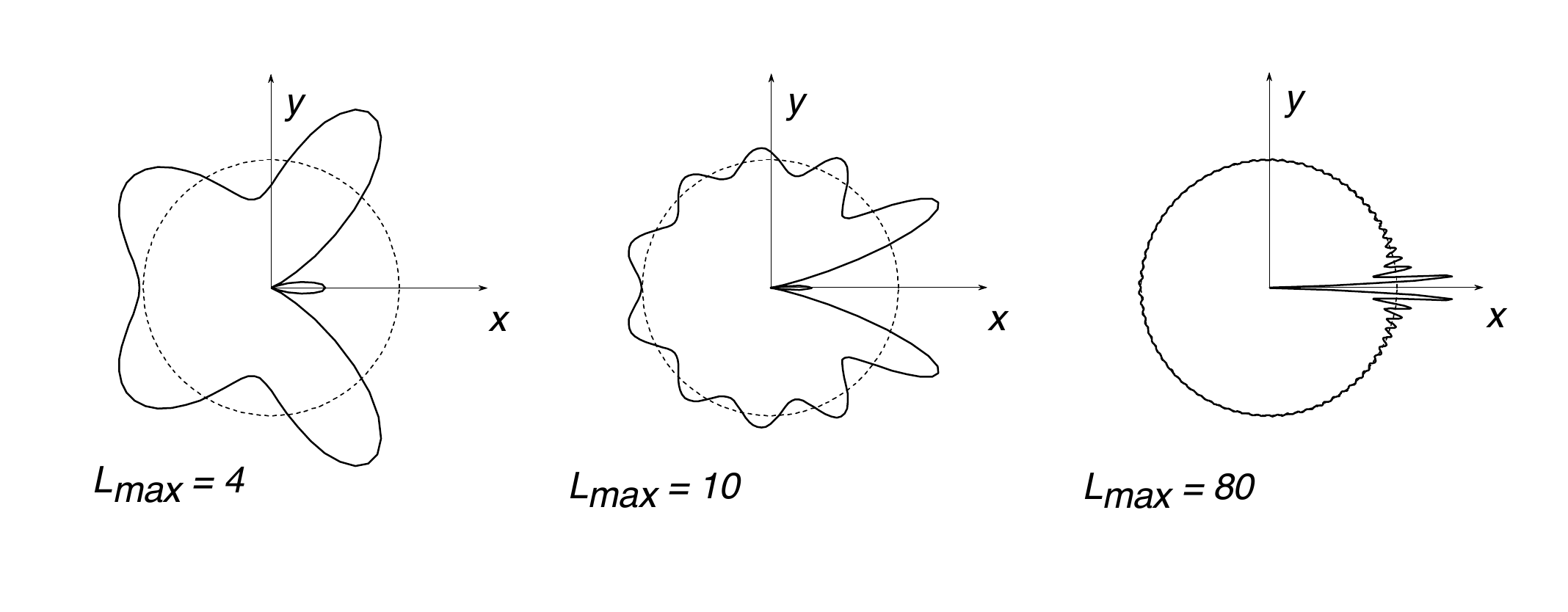}
\caption{(Back and white online) Convergence of differential cross section for various  total
  number of partial waves $L_{\rm max}$ included in the sum
  (\ref{partial_wave_scattering_amplitude}).  These panels show the cross section
  $d \sigma / d \Omega_{{\bf k}^{\prime}}$ versus scattering
  direction ${\bf k}^{\prime}$, for incident momentum ${\bf k} = {\hat x}$,
  and dipoles polarized out of the plane of the diagram.
  Solid lines are the partial wave sum, while the dashed circle is the analytic result $d \sigma / d \Omega_{{\bf k}^{\prime}} = 4/9$. }
  \label{plane_converge}
\end{figure}

The sum (\ref{partial_wave_scattering_amplitude}) is quite slow to converge.  Given that
all elements of the $T$-matrix have the same energy dependence,
{\it all} partial waves must in principle be summed.  In practice,
the result will depend on the maximum partial wave, $L_{\rm max}$,
included in a given calculation.
We show an example of this convergence in Figure \ref{plane_converge}.
In this case we choose the unsymmetrized cross section $d \sigma /d \Omega_{{\bf k}^{\prime}}$
and focus on scattering in the $x$-$y$ plane, perpendicular to the
polarization axis.  Moreover, we set the scattering length $a=0$.  In this case, we see from Eqn. (\ref{f_unsymmetrized}) that $d \sigma / d \Omega_{{\bf k}^{\prime}}$
is independent of the direction of scattering ${\bf k}^{\prime}$
in this plane, and has the value $d \sigma /  d \Omega_{{\bf k}^{\prime}} = 4/9$
in natural units.
This cross section is shown as the dashed circle in all panels of
Fig. \ref{plane_converge}.

The partial wave expansion struggles to reproduce this cross section,
as shown by the solid lines in the three panels of Figure \ref{plane_converge}.
If only four partial waves are included, the cross section does not even come close, exhibiting a shape more like a butterfly than a circle.
As more partial waves are included, the cross section conforms more
closely to the correct, circular shape.  There remain, however, large
deviations in the forward scattering direction.  These are the result
of the discontinuity in this direction, which leads to a ringing familiar from
the Gibbs phenomenon of Fourier expansion of discontinuous functions.

The discontinuity takes a concrete form within the partial wave expansion.
Setting ${\bf k}^{\prime} = {\bf k}$ in (\ref{partial_wave_scattering_amplitude}), we can
sum the series directly.  The sum over $m$, for fixed $l^{\prime}$
and $l$, can be evaluated from a spherical harmonic addition theorem \cite{BS}
\begin{eqnarray}
&& \sum_m (-1)^m Y_{l^{\prime}m}^*({\hat k})
\left( \begin{array}{ccc}
                     l^{\prime} & 2 & l \\
                     -m & 0 & m \end{array} \right)
Y_{lm}({\hat k}) \nonumber \\
&& \;\;\;\; \;\;\;\;\; = \frac{ \sqrt{(2l^{\prime}+1)(2l+1)} }{ 4 \pi }
\left( \begin{array}{ccc}
                     l^{\prime} & 2 & l \\
                     0 & 0 & 0 \end{array} \right)
C_{20}({\hat k}).
\end{eqnarray}
This reduces the directional dependence to a single reduced spherical
harmonic of the incident direction.  The forward scattering
amplitude then becomes
\begin{eqnarray}
f({\hat k},{\hat k}) = \frac {1 }{ 2 } C_{20}({\hat k})
\sum_{l^{\prime}l} i^{l^{\prime}-l} (2l^{\prime}+1)(2l+1)
\left( \begin{array}{ccc}
                     l^{\prime} & 2 & l \\
                     0 & 0 & 0 \end{array} \right)^2
\Gamma_{l^{\prime}l}
\equiv \frac {1 }{ 2 } C_{20}({\hat k}) \sum_{l^{\prime}l} A_{l^{\prime}l},
\end{eqnarray}
which defines the shorthand notation $A_{l^{\prime}l}$.
All the factors defining $A_{l^{\prime}l}$ have simple algebraic expressions
\cite{BS}, which vanish unless $l^{\prime} =l, l \pm 2$.  Inserting
these expressions, and accounting for $A_{00}=0$, we group the
terms as follows
\begin{eqnarray}
\sum_{l^{\prime}l} A_{l^{\prime}l} &=&
A_{00} + A_{02} + A_{20} + \sum_{l=1}^{\infty}
\left( A_{ll} + A_{l+2,l} + A_{l,l+2} \right) \nonumber \\
&=& -\frac{ 4 }{ 3 } + \sum_{l=1}^{\infty}
\left[ \frac{ 4(2l+1) }{ (2l-1)(2l+3) }
-\frac{ 4 }{ 2l+3 } \right].
\end{eqnarray}
As written, the second term in square brackets gives a divergent series.
However, grouping the terms as shown reduces the expression in these brackets
to $8/(2l-1)(2l+3)$, and the sum  amazingly vanishes.  The forward
scattering amplitude then becomes
\begin{eqnarray}
\label{forward_partial_wave}
f({\hat k},{\hat k}) = - \frac { 2 }{ 3 } C_{20}({\hat k})
= - \frac{ 1 }{ 3 } \left( 3\cos^2 \theta_k + 1 \right).
\end{eqnarray}
In particular, for the example shown in Fig. \ref{plane_converge}(c),
this leads to a cross section $d \sigma / d \Omega_{k^{\prime}} = 1/9$ in the forward direction,
distinct from the $4/9$ denoted by the dashed circle \footnote{The cross
section in the figure appears to vanish in the forward direction, but it
is actually zero at a small angle away from zero, and has the value $1/9$ at $\phi= 0$.}.

Based on the expression (\ref{forward_partial_wave}), the cross section, while
discontinuous, is nevertheless bounded in the forward direction.  This circumstance
is what allows us to integrate over the scattering angles and determine total
cross sections, as in Sec. II above.

We conclude that, while the partial-wave version of the Born cross
section is formally accurate, it is likely not the best way
to construct differential cross sections for elastic scattering for dipoles.  This
issue with convergence must be borne in mind when constructing more elaborate
differential scattering of molecules, with realistic interaction potentials
that also have dipolar long-range behavior.

%%%%%%%%%%%%%%%%%%%%%%%%%%%%%%%%%%%%%%%%%%%%%%%%%

\subsection{Formal convergence of the partial wave expansion}

Starting from the Born approximation (\ref{first_Born}) and substituting a partial
wave expansion for both the ingoing and outgoing waves, the
scattering amplitude reads
\begin{eqnarray}
f^{(1)}({\bf k}^{\prime},{\bf k}) = -\frac{ 1 }{ 2 \pi } \int d^3r
&& 4 \pi \sum_{l^{\prime}m^{\prime}} i^{-l^{\prime}}
Y_{l^{\prime}m^{\prime}}({\hat k}^{\prime}) j_{l^{\prime}}(k^{\prime}r)
Y_{l^{\prime}m^{\prime}}^*({\hat r}) \nonumber \\
&& \times \left( - \frac{ 2 }{ r^3 } C_{20}({\hat r}) \right) \\
&& \times
4 \pi \sum_{lm} i^l Y_{lm}^*({\hat k}) j_l(kr) Y_{lm}({\hat r}). \nonumber
\end{eqnarray}
Performing the integration over ${\bf r}$, one derives from this the partial wave series
(\ref{partial_wave_scattering_amplitude}).
Alternatively one can separate radial and angular integrals, to get
\begin{eqnarray}
f^{(1)}({\bf k}^{\prime},{\bf k}) &=& \frac{ 1 }{ \pi } (4 \pi )^2
\int_0^{\infty} r^2 dr \frac{ 1 }{ r^3 } \nonumber \\
&& \times \left[
\sum_{l^{\prime}lm}i^{-l^{\prime}+l} Y_{l^{\prime}m}^*({\hat k}^{\prime})
j_{l^{\prime}}(k^{\prime r})
C_{\l^{\prime}l}^{(m)}Y_{lm}({\hat k})j_l(kr)
\right].
\end{eqnarray}
Using a generalized version of the familiar spherical harmonic
addition theorem [Eqn. (13) of Ref. \cite{Danos65_JMP}], the quantity in square
brackets can be rewritten
\begin{eqnarray}
i^2\frac{ 1 }{ 4 \pi }\sqrt{ \frac{ 4 \pi }{ 2(2) +1 } }
 Y_{20}({\hat q}) j_2(qr),
\end{eqnarray}
where ${\bf q} = {\bf k} - {\bf k}^{\prime}$ is the familiar momentum
transfer.  The sum over partial wave quantum numbers is therefore convergent for any given
value of $qr$.  Completing the integral as in Sec. IIB. yields the
scattering amplitude (\ref{f_unsymmetrized}).  We conclude that the partial wave expansion and the 
explicit angular form of the scattering amplitude (\ref{f_unsymmetrized}) are equivalent.

%%%%%%%%%%%%%%%%%%%%%%%%%%%%%%%%%%%%%%%%

\subsection{A remark concerning the optical theorem}

The formal differential cross section for dipoles, as we have presented
it, possesses a curious property.  In general, scattering cross sections are
related to the imaginary part of the forward scattering amplitude via
the optical theorem.  For the cross sections above, however, the
value of the forward scattering amplitude is ambiguous, yet unambiguously
real-valued.

More generally, it is well-known that scattering within the first Born
approximation does not satisfy the optical theorem, but that a combination
of the first and second Born approximations does, in the form
\cite{Landau_book}
\begin{eqnarray}
\label{Born_optical}
\sigma^{(1)}({\bf k}) = \frac{ 4 \pi }{ k }
\Im [ f^{(2)}({\bf k}, {\bf k}) ],
\end{eqnarray}
where the superscripts denote the order of the Born approximation used.
We have computed the total cross section $\sigma^{(1)}$ as
a function of incident scattering direction above.  The forward
scattering amplitude, in the second-Born approximation, is
given by
\begin{eqnarray}
f^{(2)}({\bf k},{\bf k}) = -(2 \pi)^2
\int d^3 p V({\bf k},{\bf p}) G_E(p) V({\bf p},{\bf k}),
\end{eqnarray}
where
\begin{eqnarray}
G_E(p) = \frac{ 1 }{ E - p^2/2 + i \epsilon }
\end{eqnarray}
is the free-particle Green's function with outgoing boundary
conditions; and each factor $V$ is the Fourier transform of the potential,
\begin{eqnarray}
V({\bf k},{\bf p}) &=& \int d^3r \frac{ 1 }{ (2 \pi )^{3/2} }
e^{-i{\bf k} \cdot {\bf r}} V_d({\bf r}) \frac{ 1 }{ (2 \pi )^{3/2} }
e^{i{\bf p} \cdot {\bf r}}.
\end{eqnarray}
This is evaluated just as the scattering amplitude was in Sec. II,
to yield
\begin{eqnarray}
V({\bf k},{\bf p}) = \frac{ 1 }{ 3 \pi^2 } C_{20}(\theta_{q}).
\end{eqnarray}
Here, as before, $\theta_q$ is the angle between ${\bf p} - {\bf k}$
and the polarization axis.  This quantity is obviously related to
the scattering amplitude in the first Born approximation, but to
an off-shell scattering amplitude, since $k$ need not
equal $p$.

The forward scattering amplitude in (\ref{Born_optical}) is then
\begin{eqnarray}
f^{(2)}({\bf k},{\bf k}) &=& - \frac{ 4 }{ 9 \pi^2 }
\int d^3 p \frac{ |C_{20}(\theta_q)|^2 }{ E - p^2/2 + i\epsilon }
\nonumber \\
&=& - \frac{ 4 }{ 9 \pi^2 }  \int_0^{\infty} p^2 dp
\frac{ 1 }{ E - p^2/2 + i\epsilon }
\int d {\hat  p} |C_{20}(\theta_q)|^2 .
\end{eqnarray}
The integral over the direction ${\hat p}$ is easily done, and yields
$4 \pi /5$.  This leaves the radial integral
\begin{eqnarray}
f^{(2)}({\bf k},{\bf k}) = - \frac{ 16 }{ 45 \pi }
\int_0^{\infty} dp \frac{ p^2 }{ E - p^2/2 + i\epsilon }.
\end{eqnarray}
However, this integral does not converge, as the integrand
approaches a constant when $p \rightarrow \infty$.  We conclude that the second-order Born approximation is ill-defined for pure dipolar scattering, hence the optical theorem is not satisfied by our simple formulas.

This situation results from divergence at large values of the intermediate
momentum $p$, corresponding to the divergence in the
dipole potential for small $r$.  In such a divergent potential, the scattering wave function is not merely a perturbation of the free scattering wave function, as noted by Wang \cite{Wang08_NJP}.  
This situation can be remedied by, for example,
establishing a cutoff radius $b$ so that $V_d = 0$ when $r<b$.
The influence of such a cutoff was described above, in Eqn.
(\ref{radial_integral}).  The cutoff-dependent scattering amplitude is
\begin{eqnarray}
f^{(2)}({\bf k},{\bf k}) = - \frac{ 4 }{ \pi^2 }
\int d^3 p \frac{ |C_{20}(\theta_q)|^2 }{ E - p^2/2 + i \epsilon }
\left[ \frac{ \sin (qb) }{ (qb)^3 } - \frac{ \cos (qb) }{ (qb)^2 }
\right]^2.
\end{eqnarray}
Doing so complicates the integration considerably, since the magnitude
of $q = | {\bf p} - {\bf k} |$ is a function of the direction ${\hat p}$.
Nevertheless, in the limit of large $p$, $q \rightarrow p$ and
the integrand scales as $\propto p^{-6}$, which leads to a convergent integral.
This simple exercise points to a general caution: when going beyond
the first Born approximation for dipolar scattering, it will be
necessary to treat in some realistic  fashion the short-range physics
of the interaction potential.

%%%%%%%%%%%%%%%%%%%%%%%%%%%%%%%%%%%%%%%%%%%%%%

\section{Conclusions}

Reduction of the collision energy to the ultracold regime, as always, simplifies the theoretical description of scattering.  For dipolar particles, however, this simplification does {\it not} reduce to isotropy of scattering, as it does for nondipolar particles.  Rather, the differential cross section is a somewhat nontrivial function of both the incident and scattered wave vectors, of the direction of polarization of the dipoles, and of the interplay between all three directions.  To describe this scattering, we have demonstrated that it is useful to employ the direct expressions (\ref{f_unsymmetrized},\ref{f_symmetrized}), rather than slowly converging partial wave expansions.  

The isotropy that persists down to the ultracold regime has consequences for the rearrangement  of energy due to collisions.  We have shown that this anisotropy can have a profound influence on the rate of rethermalization of a gas taken out of equilibrium, since scattering at the required scattering angles can be made more or less favorable by adjusting the direction of the electric or magnetic field.  Thus even a thermal, non-quantum-degenerate gas, at ultracold temperatures, may be expected to exhibit strong anisotropy if its constituent particles are dipolar.

\acknowledgments

This work was supported by the JILA NSF Physics Frontier Center, 1125844.

\end{document}